\documentclass[a4paper,10pt,twoside]{cpc-hepnp}

\usepackage{multicol}
\usepackage{graphicx}
\usepackage{booktabs}
\usepackage{amssymb,bm,mathrsfs,bbm,amscd}
\usepackage[tbtags]{amsmath}
\usepackage{lastpage}
\usepackage{CJK}

\begin{document}

\fancyhead[c]{\small submitted to 'Chinese Physics C')} \fancyfoot[C]{\small 010201-\thepage}

 \fancyfoot[C]{\small 010201-\thepage}

\title{Calculation of the coupled betatron motion introduced by cooler's solenoids of CSRm\thanks{Supported by National Natural Science
Foundation of China (11375245) }}

\author{%
      TANG Mei-Tang （汤梅堂）$^{1,2;1)}$\email{tangmt@impcas.ac.cn}%
\quad YANG Xiao-Dong （杨晓东）$^{1}$
\quad MAO Li-Jun （冒立军）$^{1}$\\
\quad LI  Jie （李杰）$^{1}$
\quad MA Xiao-Ming （马晓明）$^{1}$
\quad YAN Tai-Lai （晏太来）$^{1}$\\
\quad ZHENG Wen-Heng （郑文亨）$^{1,2}$
\quad ZHAO He（赵贺）$^{1,2}$
\quad WU Bo （吴波）$^{1,2}$\\
\quad WANG Geng （王耿）$^{1,2}$
\quad RUAN Shuang （阮爽）$^{1,2}$
\quad SHA Xiao-Ping （沙小平）$^{1}$\\
\quad YANG Jian-cheng （杨建成）$^{1}$
\quad YUAN You-Jin （原有进）$^{1}$
\quad XIA Jia-Wen （夏佳文）$^{1}$
}
\maketitle

\address{%
$^1$ Institute of Modern Physics, Chinese Academy of Sciences, Lanzhou 730000, People's Republic of China\\
$^2$ University of Chinese Academy of Sciences, Beijing 100049, People's Republic of China\\
}

\begin{abstract}
Several solenoids are usually installed in electron cooler device to guide the motion of the electron beam in the cooler. However, the solenoids also have influence to the ion beam in the cooler storage ring. The transverse motion of the ion beam in storage ring will become coupled, if the solenoids installed in the electron cooler are not compensated perfectly. In this paper, the coupled transverse motion due to the uncompensated cooler's solenoids of CSRm (The main storage ring in the IMP, Lan Zhou, China) is studied, and the coupled beam's envelopes are calculated by a new method.
\end{abstract}



\footnotetext[0]{\hspace*{-3mm}\raisebox{0.3ex}{$\scriptstyle\copyright$}2013
Chinese Physical Society and the Institute of High Energy Physics
of the Chinese Academy of Sciences and the Institute
of Modern Physics of the Chinese Academy of Sciences and IOP Publishing Ltd}%

\begin{multicols}{2}

\section{Introduction}

An electron cooler device named EC-35 is installed in the CSRm to improve the quality of the ion beam in the storage ring[1]. As Figure 1 shows, there are several solenoids installed in the electron cooler, and revolution ion beam passes through the whole main solenoid and parts of the two toroid every turn. If the solenoids are not compensated, it will make the beam's betatron motion of horizontal and vertical directions become coupled. Usually, two compensation solenoids installed outside the electron cooler are used to decouple the beam.

\begin{center}
\includegraphics[width=6.5cm]{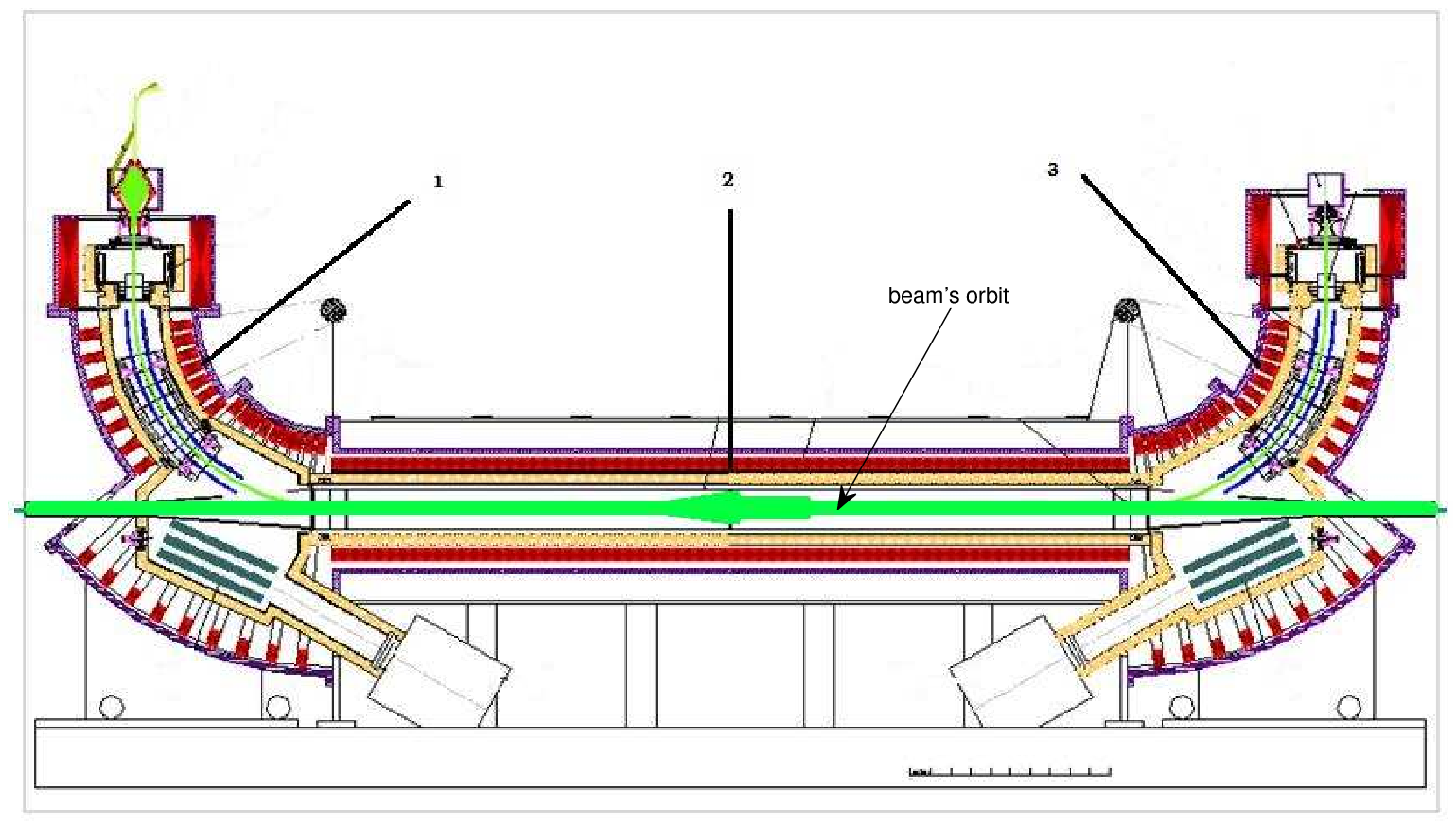}
\figcaption{\label{fig1} The general arrangement of the EC-35, 1, 3 are the toroids, 2 is the main solenoid.}
\end{center}

In this paper the transverse motion of ion beam in CSRm will be studied when the compensation solenoids are not used or are turned off suddenly. To research the coupled motion two methods are used: the tracking method and the Ripken's analytical method.

In Section 2 the simulation results of the coupled betatron motion will be presented, in Section 3 the Ripken's method is used to parameterize two dimensional coupled betatron motion of CSRm, and the boundary functions in phase plane for coupled single particle and beam are obtained. In Section 4 a new method used to calculate the coupled beam envelopes are introduced and the calculating results are presented. In Section 5 the conclusions are summarized.

\section{Simulation of the coupled betatron motion}

The 4*4 transmission matrixes of every element in the ring CSRm are used to simulate the coupled betatron motion. Particularly, transmission matrixes of the solenoid are presented by Equation (1), when the direction of the longitudinal magnetic fields in the solenoid is opposite to the particles' motion direction[2](The situation in the CSRm):

\begin{equation}\label{1}
M = \left( {\begin{array}{*{20}{c}}
{\frac{{1 + C}}{2}}&{\frac{S}{K}}&{ - \frac{S}{2}}&{ - \frac{{1 - C}}{K}}\\
{ - \frac{{KS}}{4}}&{\frac{{1 + C}}{2}}&{\frac{{K(1 - C)}}{4}}&{ - \frac{S}{2}}\\
{\frac{S}{2}}&{\frac{{1 - C}}{K}}&{\frac{{1 + C}}{2}}&{\frac{S}{K}}\\
{ - \frac{{K(1 - C)}}{4}}&{\frac{S}{2}}&{ - \frac{{KS}}{4}}&{\frac{{1 + C}}{2}}
\end{array}} \right).\
\end{equation}
Where $K=B_z/G, S=sin(KL), C=cos(KL), B_z$ is the value of longitudinal fields of solenoid, L is the length of solenoid, G is the magnetic rigidity of ion particle.

\begin{center}
\includegraphics[width=6.5cm]{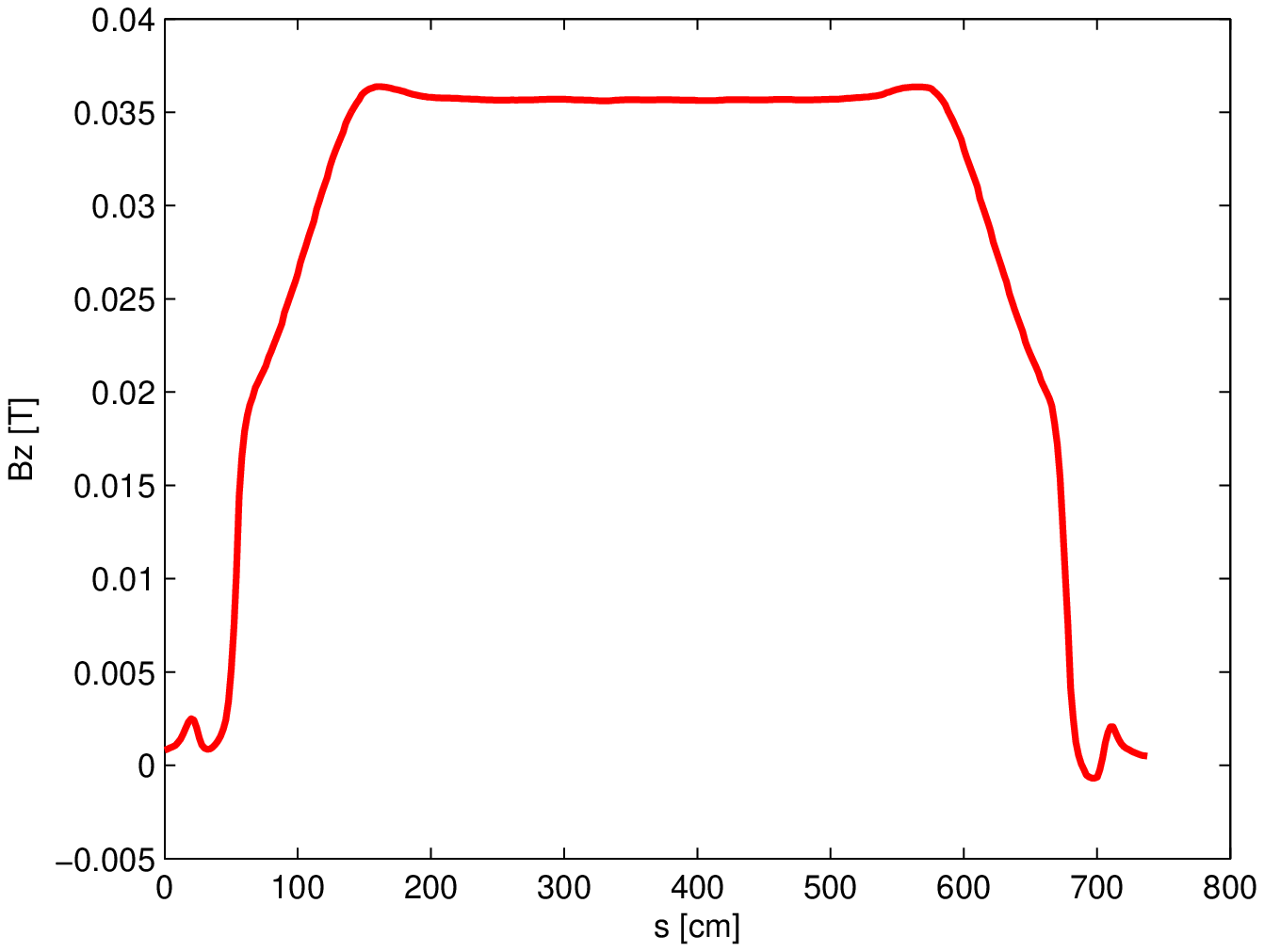}
\figcaption{\label{fig2} The longitudinal component of magnet fields in the cooler section of CSRm}
\end{center}

In our case, the longitudinal magnetic fields in the cooler section of CSRm are not uniform, the value changes along longitudinal position, as Figure 2 shows. So we can not treat the whole cooler section as a solenoid. In order to obtain more accurate results, the transmission matrixes of the solenoids in cooler section are obtained by following ways:

Divide the cooler section into several parts, each part acting as a independent solenoid has a 4*4 transmission matrixes. By averaging the magnetic data of each part, the $B_z $ can be obtained, then using equation (1) the transmission matrixes of the each part can be calculated out.

7MeV/u $^{12}C^{6+}$ is chosen as example particle in simulation. Assuming that the initial particles matching the uncoupled twiss function at the starting point, and the emittances are equal to the acceptances of the CSRm. The detail parameters for initial particles are shown in Table 1[3].
\begin{center}
\tabcaption{ \label{tab1}  Parameters used in the calculations.}
\footnotesize
\begin{tabular*}{80mm}{@{\extracolsep{\fill}}c|c}
\toprule Particle & 7MeV/u $^{12}C^{6+}$  \\
\hline
Horizontal acceptance $A_x$  & $200\pi mmmrad$\\
\hline
Vertical acceptance $A_y $   & $30\pi mmmrad$  \\
\hline
Horizontal $\beta$ function  &$\beta_x$ = $4.68m$\\
\hline
Vertical $\beta$ function  &$\beta_y$ = $30.17m$\\
\hline
Horizontal $\alpha$ function &  $\alpha _x $= 0\\
\hline
Vertical $\alpha$ function &  $\alpha _y $= 0\\
\bottomrule
\end{tabular*}%
\end{center}

$1>$ Single particle's motion

A particle is generated randomly in (x, x', y, y') phase plane, then use transmission matrixes to track the particle turn by turn. At one fixed watching point, the particle's projection in phase plane can be obtained turn by turn, as Figure 3,4 show.

\begin{center}
\includegraphics[width=6.5cm]{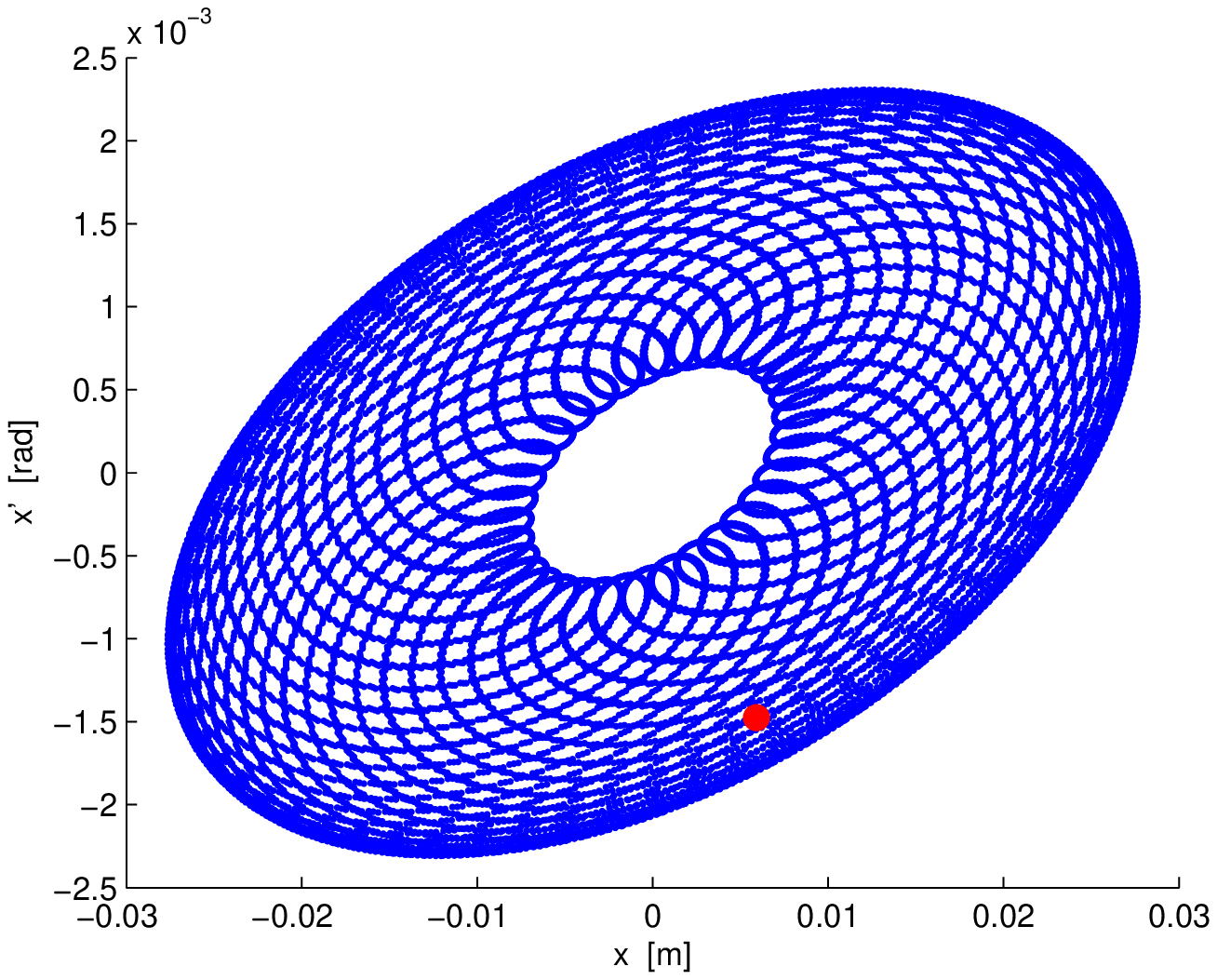}
\figcaption{\label{fig3} The single particle's projection in (x,x') phase plane for 20000 turns in one watching point of the ring. Red point is the initial position of the particle.}
\end{center}

Form the Figure 3, one can find that: with coupling, a single particle's projection for 20000 turns is different from the uncoupled case, the coupled particle's projection for 20000 turns at watching point no longer be a ellipse, instead it's a torus.

\begin{center}
\includegraphics[width=6.5cm]{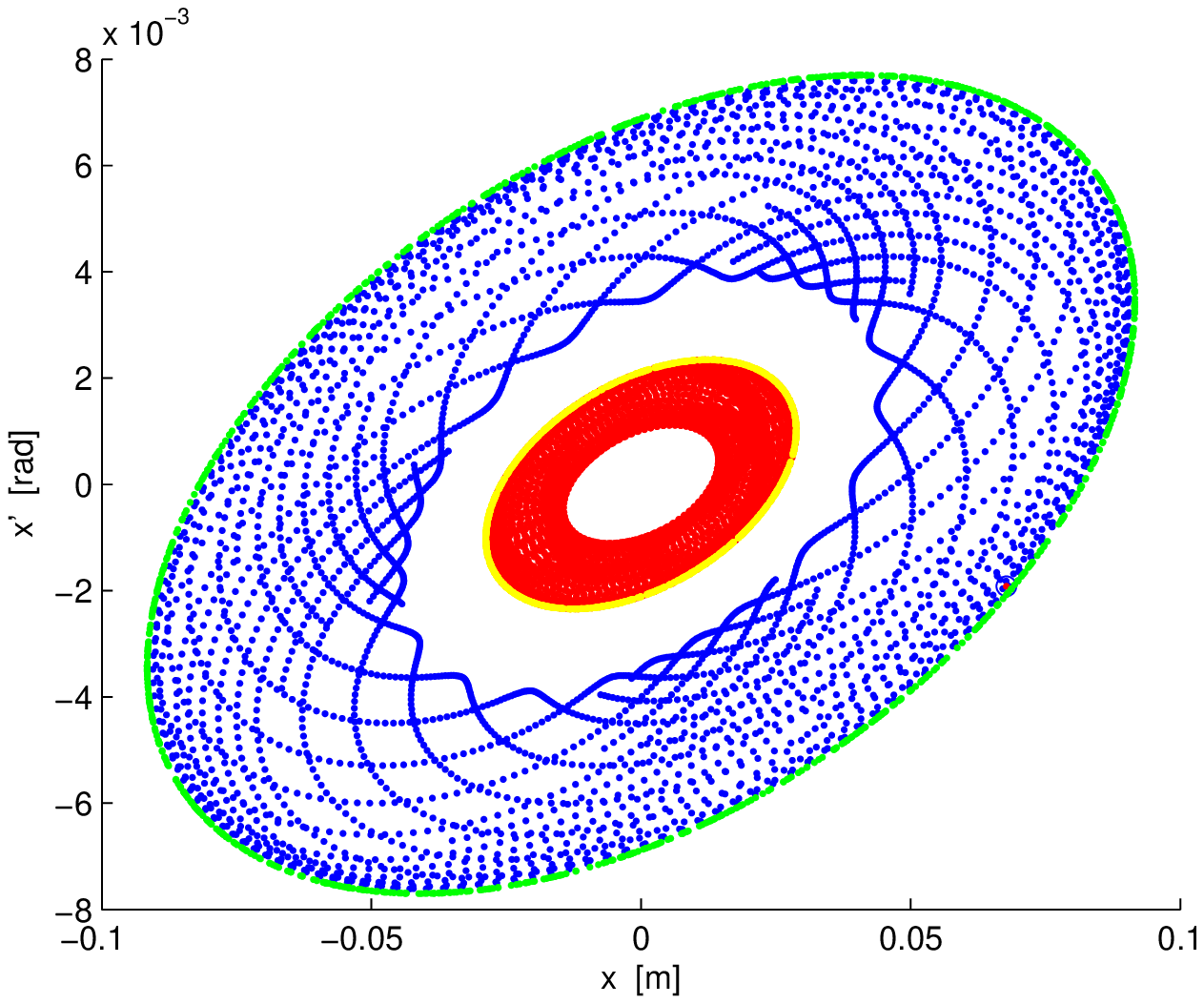}
\figcaption{\label{fig4} Two particles' projection in (x,x') phase plane for 2000 turns in a watching point of the ring. The red point are for one particle, the blue points are for another.}
\end{center}

Further simulation indicates that: $1.$ For a same particle, the shape of the projection torus is different at different watching points of the ring. 2. For different initial particles, the shapes of the projection torus's boundary are same at same watching poind, but the sizes are different (Figure 4 shows).  3. For a particle the boundary of the projection torus at every watching point will become fixed after several turns, the particle's footprint will not beyond the boundary with the increase of the turn.

$2>$ Beam's motion

To simulate the beam's motion in coupled situation, 1000 particles are generated with the parameters shown in Table 1. Then use the transmission matrixes to track the particles turn by turn. The 1000 particles' footprints in phase plane at one watching point for 2000 turns are shown in Figure 5.

\begin{center}
\includegraphics[width=6.5cm]{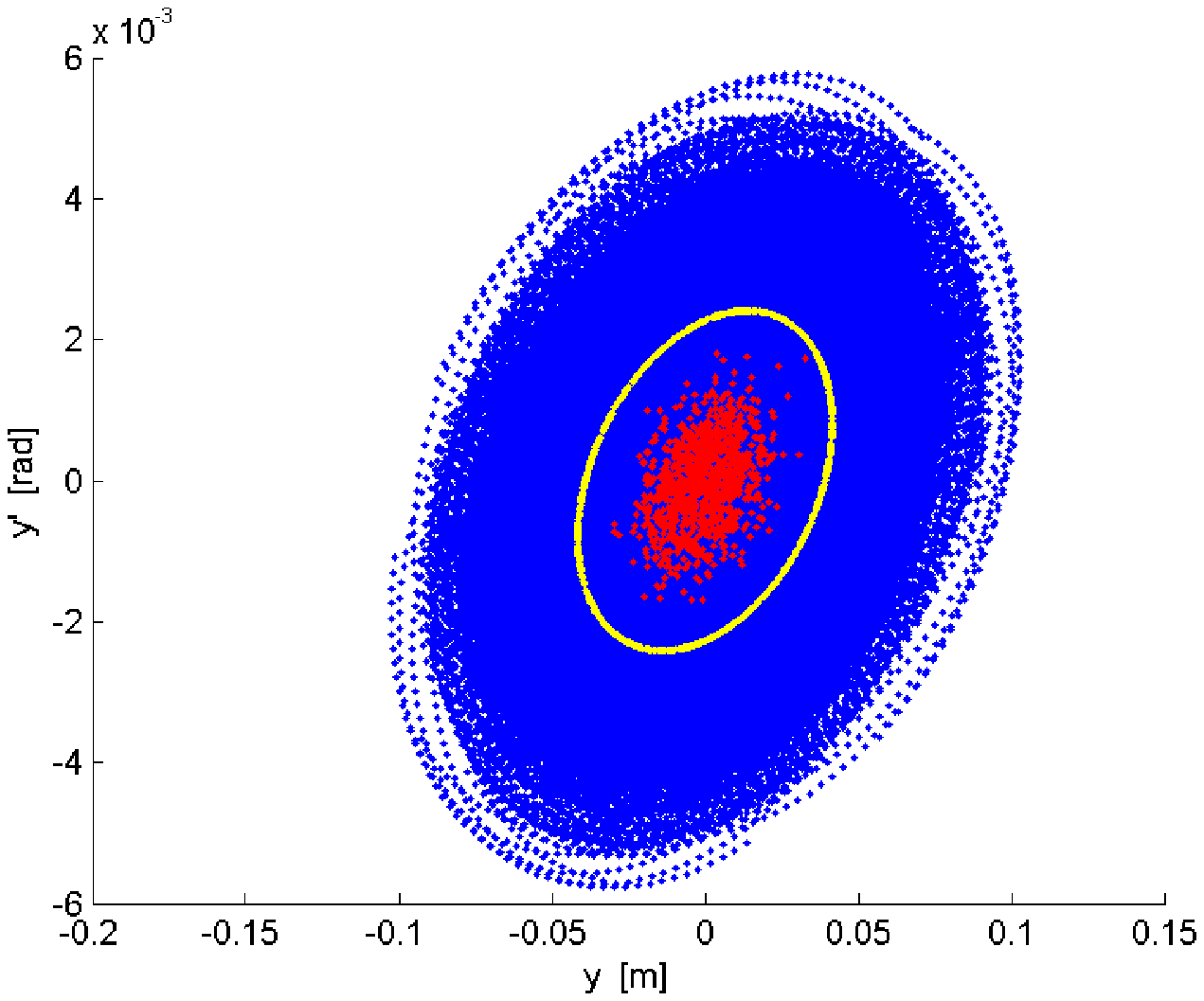}
\figcaption{\label{fig5} Beam's projection in (y,y') phase plane for 2000 turns in a watching point of the ring. The yellow ellipse is the initial matched emittance circle. Red points are the initial beam.}
\end{center}

From Figure 5 one can find that: with coupling, the beam's projection for 2000 turns in phase plane is beyond the initial emittance circle. Same as the uncoupled case，the boundary of the projection is also a ellipse.

Further simulation indicates that: 1. The boundary of the beam's projection in phase plane is different ellipse at different watching points. 2. For certain watching point the boundary ellipse is fixed, and not changes with the revolution of the beam. In other words, the envelopes along the ring for certain initial beam are certain, just like the uncoupled case.

$3>$ The variation of emittance

The variation of emittances in horizontal and in vertical directions are also calculated turn by turn, the result is shown in Figure 6.

\begin{center}
\includegraphics[width=6.5cm]{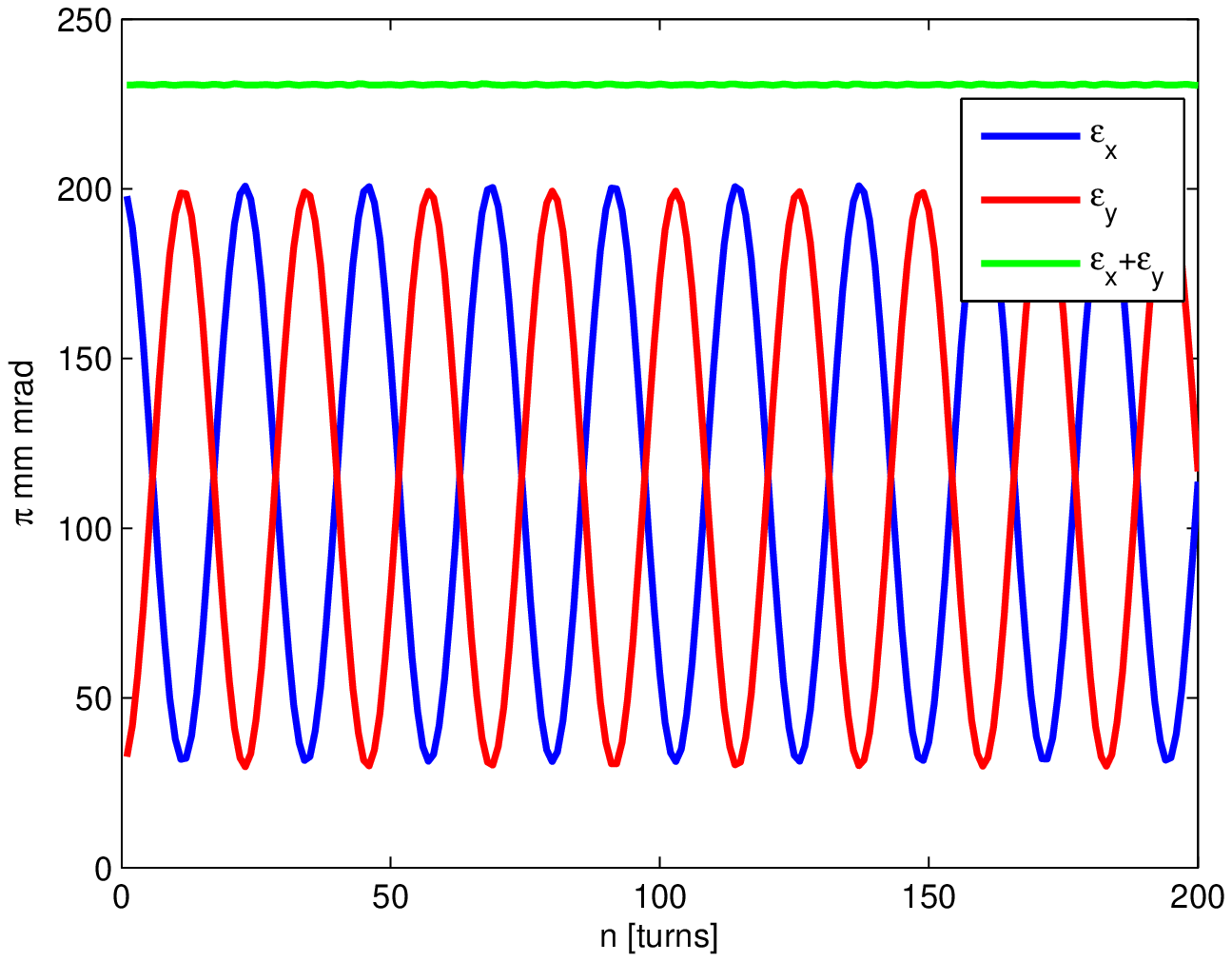}
\figcaption{\label{fig6} The emittances' changes turn by turn.}
\end{center}

From Figure 6, it can be found that the emittances in horizontal and vertical directions are not longer constants, however, the sum of the two emittances is a invariant.

From above simulation results, it become clear that the coupled betatron motion of beam is quite different from the uncoupled case. In coupled case, the beam emittances is no longer invariant. Although the beam envelope is existing, the equation $r= \sqrt {{\rm{\varepsilon }}{\rm{\beta }}}$ parameterized by Courant-Snyder parameters is no longer available to calculate the couple beam's envelope. To calculate the coupled beam envelope and represent the coupled beam motion of CSRm, the method parameterize two dimensional linearly coupled betatron motion should be introduced.

\section{The coupled motion's parameterizations}
There are several methods usually used to parameterize the coupled betatron motion. Ripken's approach and the approach obtained by Edwards and Teng are most well-known of them[4]. Although the Ripken's approach is more complicated, it can give straightforward expressions of the particle's coordinates in the laboratory frame[5]. So Ripken's approach is chosen to parameterize our coupling situation.
In Ripken's theory a particle's trajectory can be expressed by several lattice functions as[5]:
\begin{equation}\label{2}
\begin{array}{l}
\left( {\begin{array}{*{20}{c}}
{\begin{array}{*{20}{c}}
{\rm{x}}\\
{{\rm{x'}}}
\end{array}}\\
{\begin{array}{*{20}{c}}
{\rm{y}}\\
{{\rm{y'}}}
\end{array}}
\end{array}} \right) = \\
\left( {\begin{array}{*{20}{c}}
{\sqrt {{{\rm{\varepsilon }}_1}{{\rm{\beta }}_{{\rm{x}}1}}} \cos \left( {{{\rm{\Phi }}_{{\rm{x}}1}} + {{\rm{\Phi }}_1}} \right) + \sqrt {{{\rm{\varepsilon }}_2}{{\rm{\beta }}_{{\rm{x}}2}}} {\rm{cos}}\left( {{{\rm{\Phi }}_{{\rm{x}}2}} + {{\rm{\Phi }}_2}} \right)}\\
{\sqrt {{{\rm{\varepsilon }}_1}{{\rm{\gamma }}_{{\rm{x}}1}}} \cos \left( {{\phi _{{\rm{x}}1}} + {{\rm{\Phi }}_1}} \right) + \sqrt {{{\rm{\varepsilon }}_2}{{\rm{\gamma }}_{{\rm{x}}2}}} {\rm{cos}}\left( {{\phi _{{\rm{x}}2}} + {{\rm{\Phi }}_2}} \right)}\\
{\sqrt {{{\rm{\varepsilon }}_1}{{\rm{\beta }}_{{\rm{y}}1}}} \cos \left( {{{\rm{\Phi }}_{{\rm{y}}1}} + {{\rm{\Phi }}_1}} \right) + \sqrt {{{\rm{\varepsilon }}_2}{{\rm{\beta }}_{{\rm{y}}2}}} {\rm{cos}}({{\rm{\Phi }}_{{\rm{y}}2}} + {{\rm{\Phi }}_2})}\\
{\sqrt {{{\rm{\varepsilon }}_1}{{\rm{\gamma }}_{{\rm{y}}1}}} \cos \left( {{\phi _{{\rm{y}}1}} + {{\rm{\Phi }}_1}} \right) + \sqrt {{{\rm{\varepsilon }}_2}{{\rm{\gamma }}_{{\rm{y}}2}}} {\rm{cos}}\left( {{\phi _{{\rm{y}}2}} + {{\rm{\Phi }}_2}} \right)}
\end{array}} \right)
\end{array}.\
\end{equation}

Where $\epsilon_1$,$\epsilon_2$ are constants for a particle. $\beta_{x1}$, $\beta_{x2}$,\\ $\gamma_{x1}$, $\gamma_{x2}$, $\beta_{y1}$, $\beta_{y2}$, $\phi_{x1}$, $\phi_{x2}$, $\Phi_{x1}$, $\Phi_{x2}$, $\phi_{y1}$, $\phi_{y2}$, $\Phi_{y1}$, $\Phi_{y2}$ are Ripken's lattice functions and phase advances. $\phi_1$, $\phi_2$ are initial phases of a particle. These lattice functions at different position of ring have different values, and they are periodic functions with the period equal to the circumference of the ring, just like the Courant-Snyder parameters in uncoupled case.\\
\\
$1>$ Ripken's lattice functions for CSRm
\\

Ripken's lattice functions at different position can be calculated from the eigenvectors of the whole ring transmission matrixes[5]; Figure 7, 8 show the Ripken's $\beta$ functions $\beta_{x1}$, $\beta_{x2}$, $\beta_{y1}$, $\beta_{y2}$, of the ring CSRm and the Courant-Snyder parameters $\beta_{x}$, $\beta_{y}$.

\begin{center}
\includegraphics[width=6.5cm]{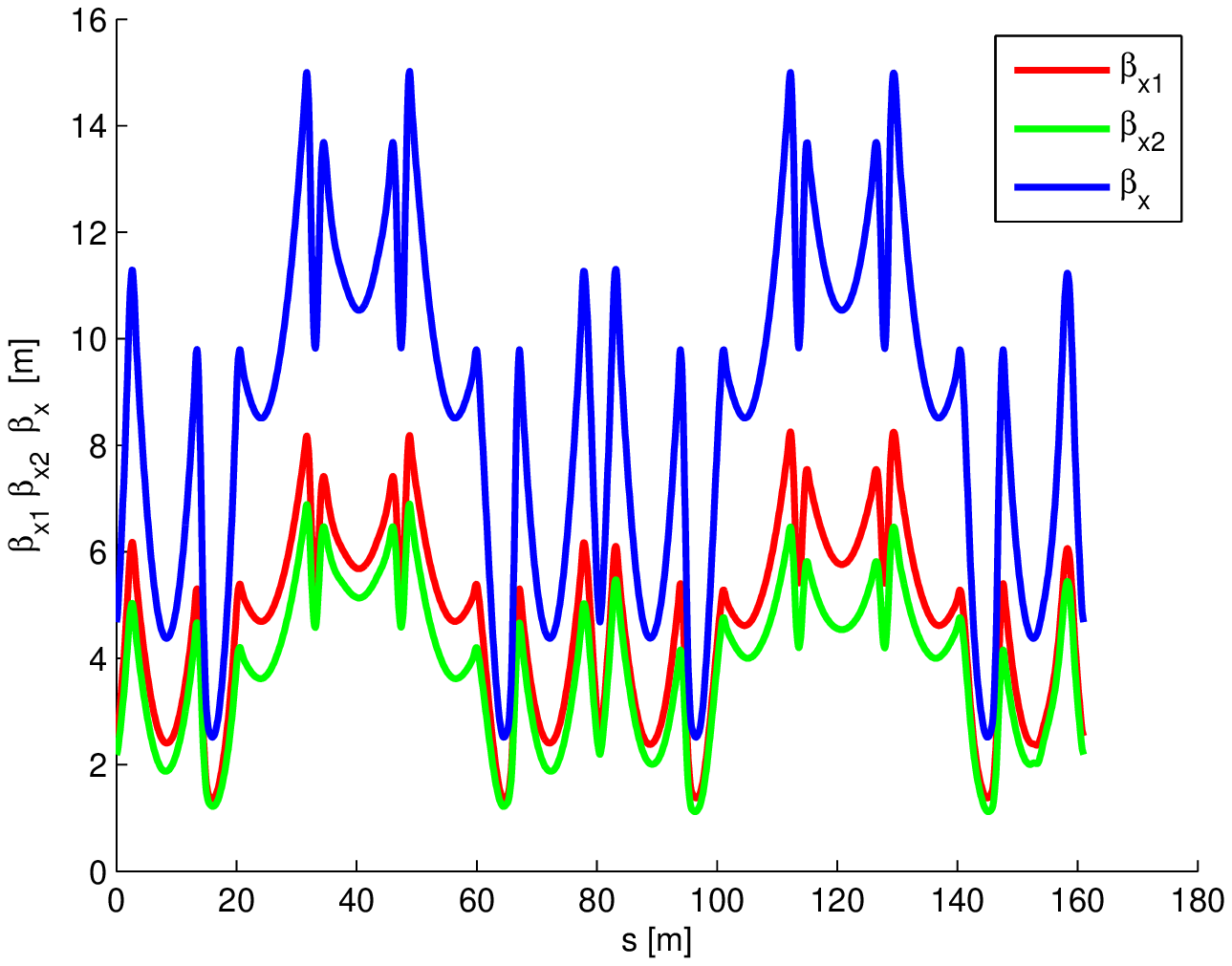}
\figcaption{\label{fig7} Ripken's $\beta$ functions with coupling and the Courant-Snyder $\beta$ function without coupling in x direction for CSRm.}
\end{center}

\begin{center}
\includegraphics[width=6.5cm]{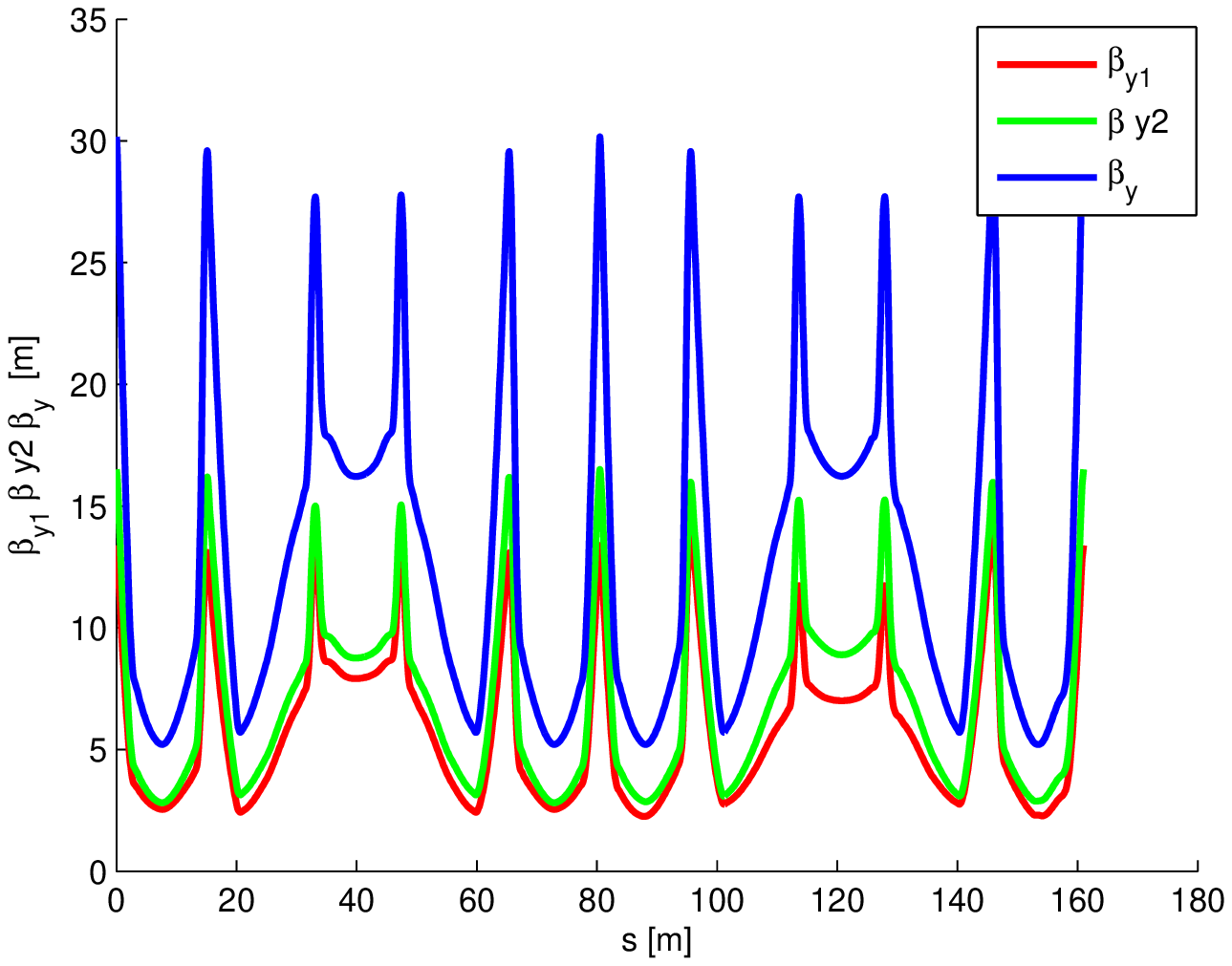}
\figcaption{\label{fig8} Ripken's $\beta$ functions with coupling and the Courant-Snyder $\beta$ function without coupling in y direction for CSRm.}
\end{center}

It can be found that$\beta_{x1}, \beta_{x2}$ have the same shape with Courant-Snyder parameters$\beta_x$, and $\beta_{y1}, \beta_{y2}$ have the same shape with$\beta_y$.\\
\\
$2>$ The boundary function of coupled beam's projection
\\

After obtained the Ripken's lattice functions of CSRm, now the Equation (2) can be used to explain beam's coupled betatron motion in CSRm. From Equation (2) it can be found that a particle's projection in (x, x') phase plane or in (y, y') phase plane is separately determined by two modes, the mode 1  parameterized by the Ripken's lattice functions with subscript 1, and mode 2 with subscript 2. Each independent mode's projection in phase plane with different phase $\phi_1$, or $\phi_2$  is ellipse.

Hence a particle's footprints on phase plane at one fixed watching point are determined by the two ellipses, as Figure 9 shows.

\begin{center}
\includegraphics[width=6.5cm]{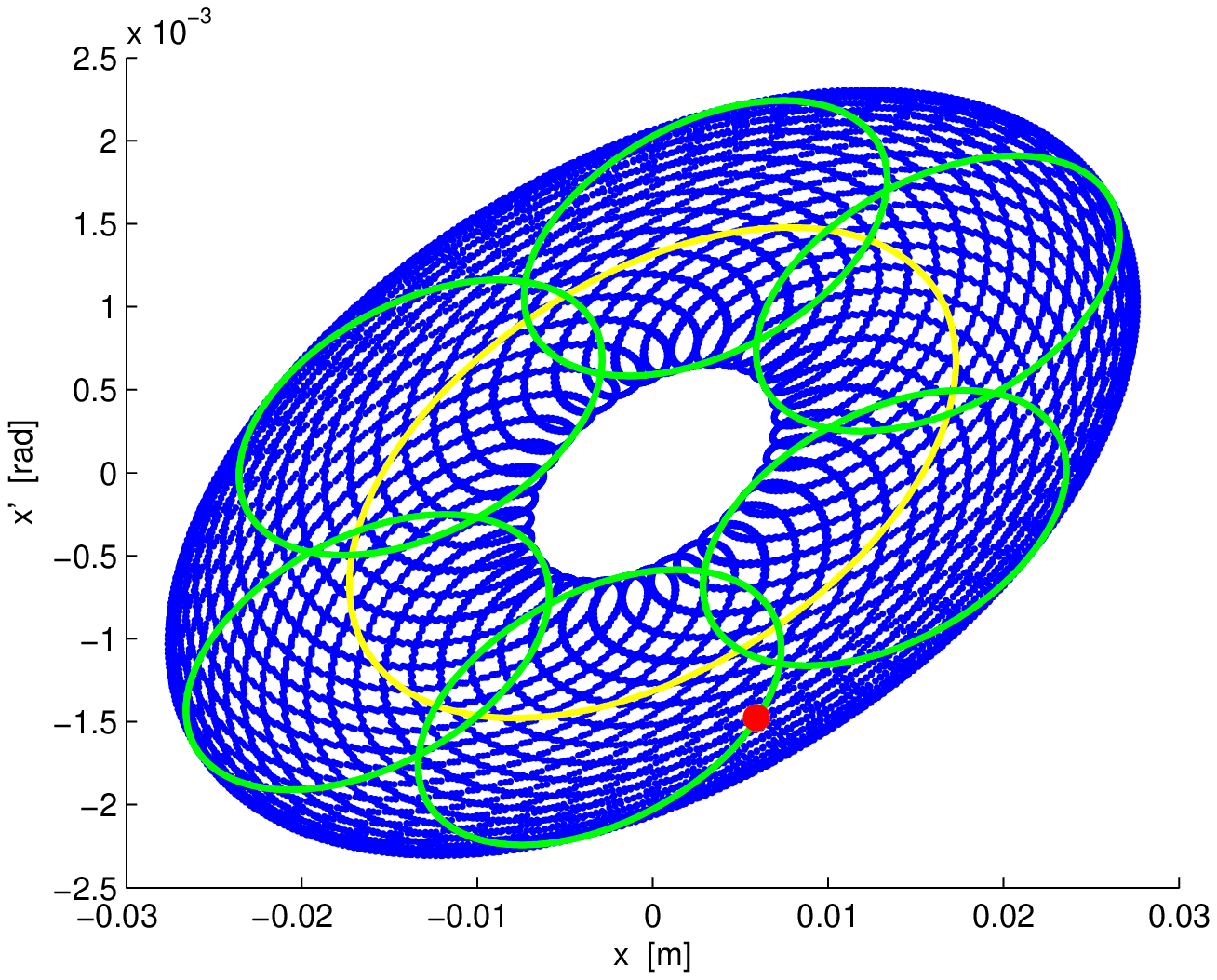}
\figcaption{\label{fig9} A particle's projection for 20000 turns in one point of the ring(blue point). The yellow and green cycles present the projections of two modes.}
\end{center}

 In simulation one find that the boundary of a particle's projection in a watching point is fixed, now the function of the boundary ellipse can be obtained from Equation (2). The boundary ellipse should pass through the point ($X_{max}, X'$), and the maximum vertical coordinate of the boundary ellipse is equal to $X'_{max}$. $X_{max}, X', X'_{max}$ are determined by Equation (3), (4), (5).

\begin{equation}\label{3}
  {{\rm{X}}_{{\rm{max}}}} = \sqrt {{{\rm{\varepsilon }}_1}{{\rm{\beta }}_{{\rm{x}}1}}}  + \sqrt {{{\rm{\varepsilon }}_2}{{\rm{\beta }}_{{\rm{x}}2}}.}
\end{equation}

\begin{equation}\label{4}
  X{\rm{'}} = \sqrt {{{\rm{\varepsilon }}_1}/{{\rm{\beta }}_{{\rm{x}}1}}} {\rm{*}}{{\rm{\alpha }}_{{\rm{x}}1}} + \sqrt {{{\rm{\varepsilon }}_2}/{{\rm{\beta }}_{{\rm{x}}2}}} {{\rm{\alpha }}_{{\rm{x}}2}}.
\end{equation}

\begin{equation}\label{5}
  {{\rm{X'}}_{{\rm{max}}}} = \sqrt {{{\rm{\varepsilon }}_1}{{\rm{\gamma }}_{{\rm{x}}1}}}  + \sqrt {{{\rm{\varepsilon }}_2}{{\rm{\gamma }}_{{\rm{x}}2}}}.
\end{equation}
Where $\alpha _{x1}, \alpha _{x2}$  are Ripken's lattice functions, which can be calculated by whole ring transmission matrixes[5].
Use above conditions, the boundary ellipse's function can be obtained:
\begin{equation}\label{6}
  \left\{ {\begin{array}{*{20}{c}}
{x = {X_{\max }}\cos (t)}\\
{x' =  - X'\cos (t) - \sqrt {X{'_{\max }}^2 - X{'^2}} \sin (t)}
\end{array}} \right..\
\end{equation}
Where t is free parameter.
Equations (3),(4), (5), (6)  are also suit for the (y, y') phase plane, in that case one only need to replace all x in above equations by y.

Figure 10 shows a particle's projection for 2000 turns and the corresponding boundary ellipse (The yellow ellipse) calculated by Equations (6).
\begin{center}
\includegraphics[width=6.5cm]{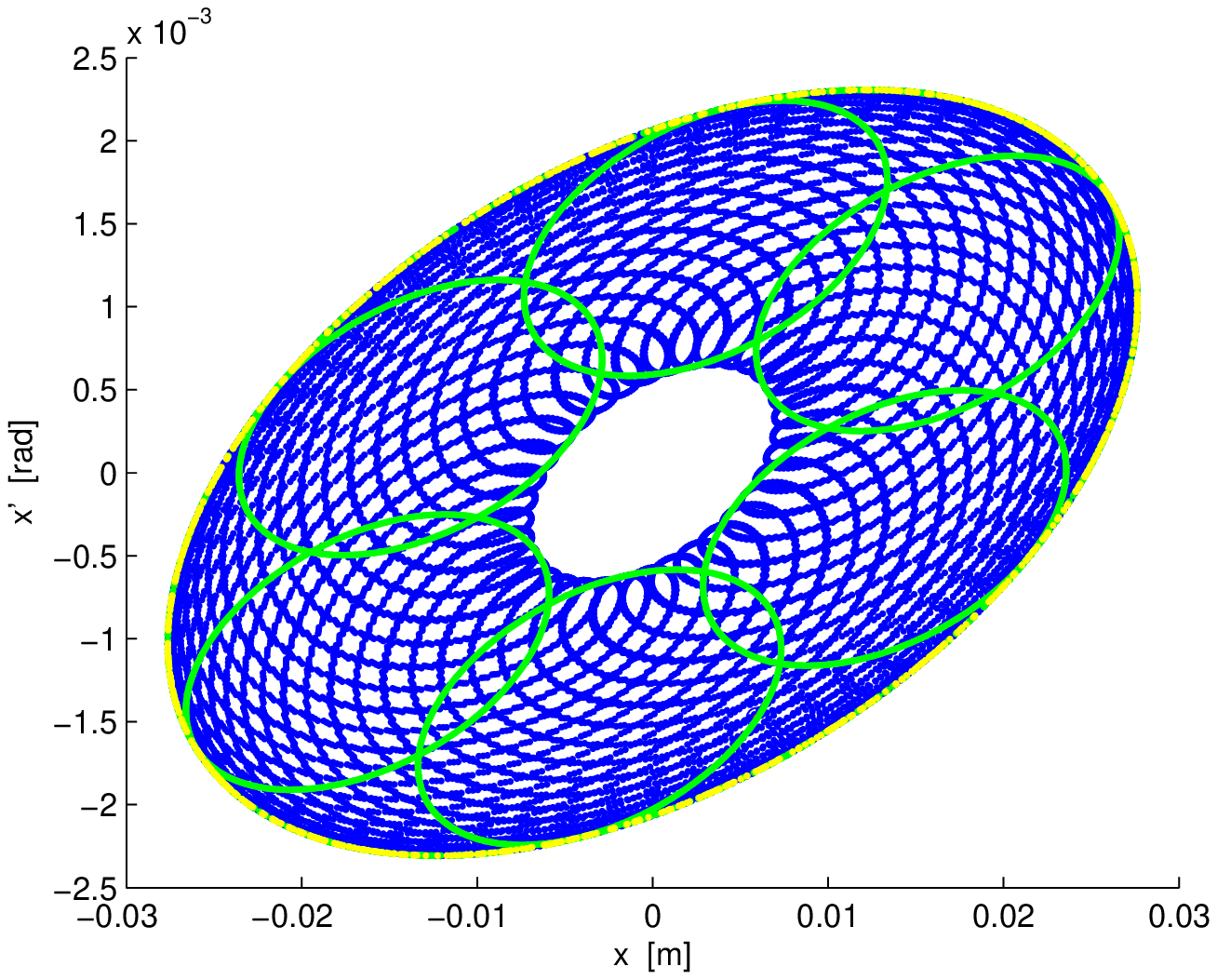}
\figcaption{\label{fig10} A particle's footprint (blue point) and its boundary ellipse (yellow one) in phase plane.}
\end{center}

Form Figure 10 it can be found that the single particle's footprint is surrounded by the boundary ellipse perfectly.

From Equation (2), the boundary ellipse function for beam can also be obtained:
\begin{equation}\label{7}
  \left\{ {\begin{array}{*{20}{c}}
{x = {Xb_{\max }}\cos (t)}\\
{x' =  - Xb'\cos (t) - \sqrt {Xb{'_{\max }}^2 - Xb{'^2}} \sin (t)}
\end{array}} \right..\
\end{equation}

\begin{equation}\label{6}
  {{\rm{Xb}}_{{\rm{max}}}} = \sqrt {{{{\rm{\mathord{\buildrel{\lower3pt\hbox{$\scriptscriptstyle\frown$}}
\over \varepsilon } }}}_1}{{\rm{\beta }}_{{\rm{x}}1}}}  + \sqrt {{{{\rm{\mathord{\buildrel{\lower3pt\hbox{$\scriptscriptstyle\frown$}}
\over \varepsilon } }}}_2}{{\rm{\beta }}_{{\rm{x}}2}}},
\end{equation}
\begin{equation}\label{7}
  {\rm{Xb'}} = \sqrt {{\rm{\;}}{{\frac{{{{{\rm{\mathord{\buildrel{\lower3pt\hbox{$\scriptscriptstyle\frown$}}
\over \varepsilon } }}}_1}}}{{{{\rm{\beta }}_{{\rm{x}}1}}}}}_{\rm{\;}}}} {\rm{*}}{{\rm{\alpha }}_{{\rm{x}}1}} + \sqrt {\frac{{{{{\rm{\mathord{\buildrel{\lower3pt\hbox{$\scriptscriptstyle\frown$}}
\over \varepsilon } }}}_2}}}{{{{\rm{\beta }}_{{\rm{x}}2}}}}} {{\rm{\alpha }}_{{\rm{x}}2}},
\end{equation}
\begin{equation}\label{8}
  {{\rm{Xb'}}_{{\rm{max}}}} = \sqrt {{{{\rm{\mathord{\buildrel{\lower3pt\hbox{$\scriptscriptstyle\frown$}}
\over \varepsilon } }}}_1}{{\rm{\gamma }}_{{\rm{x}}1}}}  + \sqrt {{{{\rm{\mathord{\buildrel{\lower3pt\hbox{$\scriptscriptstyle\frown$}}
\over \varepsilon } }}}_2}{{\rm{\gamma }}_{{\rm{x}}2}}}. \
\end{equation}

Where the ${{{{\rm{\mathord{\buildrel{\lower3pt\hbox{$\scriptscriptstyle\frown$}}
\over \varepsilon } }}}_1}}$ and ${{{{\rm{\mathord{\buildrel{\lower3pt\hbox{$\scriptscriptstyle\frown$}}
\over \varepsilon } }}}_2}}$ are the values of $\epsilon_1$, $\epsilon_2$, related to the beam's coordinates (x, x', y, y) by Equation (2), with these particular values ${{{{\rm{\mathord{\buildrel{\lower3pt\hbox{$\scriptscriptstyle\frown$}}
\over \varepsilon } }}}_1}}$ and ${{{{\rm{\mathord{\buildrel{\lower3pt\hbox{$\scriptscriptstyle\frown$}}
\over \varepsilon } }}}_2}}$   the function $\sqrt {{{\rm{\varepsilon }}_1}{{\rm{\beta }}_{{\rm{x}}1}}}  + \sqrt {{{\rm{\varepsilon }}_2}{{\rm{\beta }}_{{\rm{x}}2}}} $ should be maximized.

Above equations are also suit for the (y, y') phase plane, for y direction one just need to replace all x in above equations by y.

Figure 11, 12 show projection of beam and the boundary ellipses are calculated by Equation (7). Figure 12 is for x direction, and Figure11 is for y direction.

\begin{center}
\includegraphics[width=6.5cm]{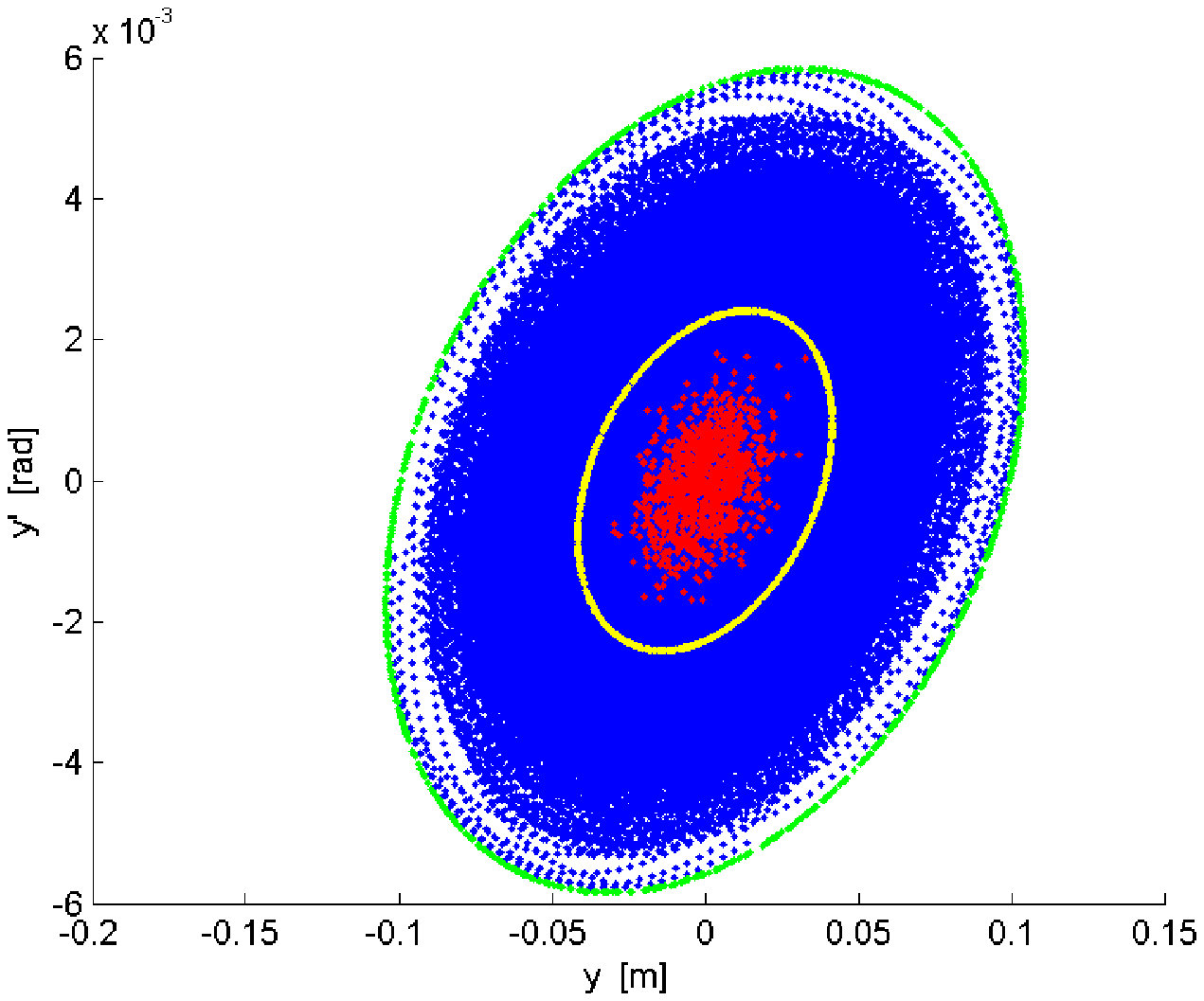}
\figcaption{\label{fig11} Projection of 1000 particles for 2000 turns in y direction. The yellow ellipse is the matched emittance cycle for initial beam. Red points are the initial beam. The green ellipse is the boundary cycle.}
\end{center}

\begin{center}
\includegraphics[width=6.50cm]{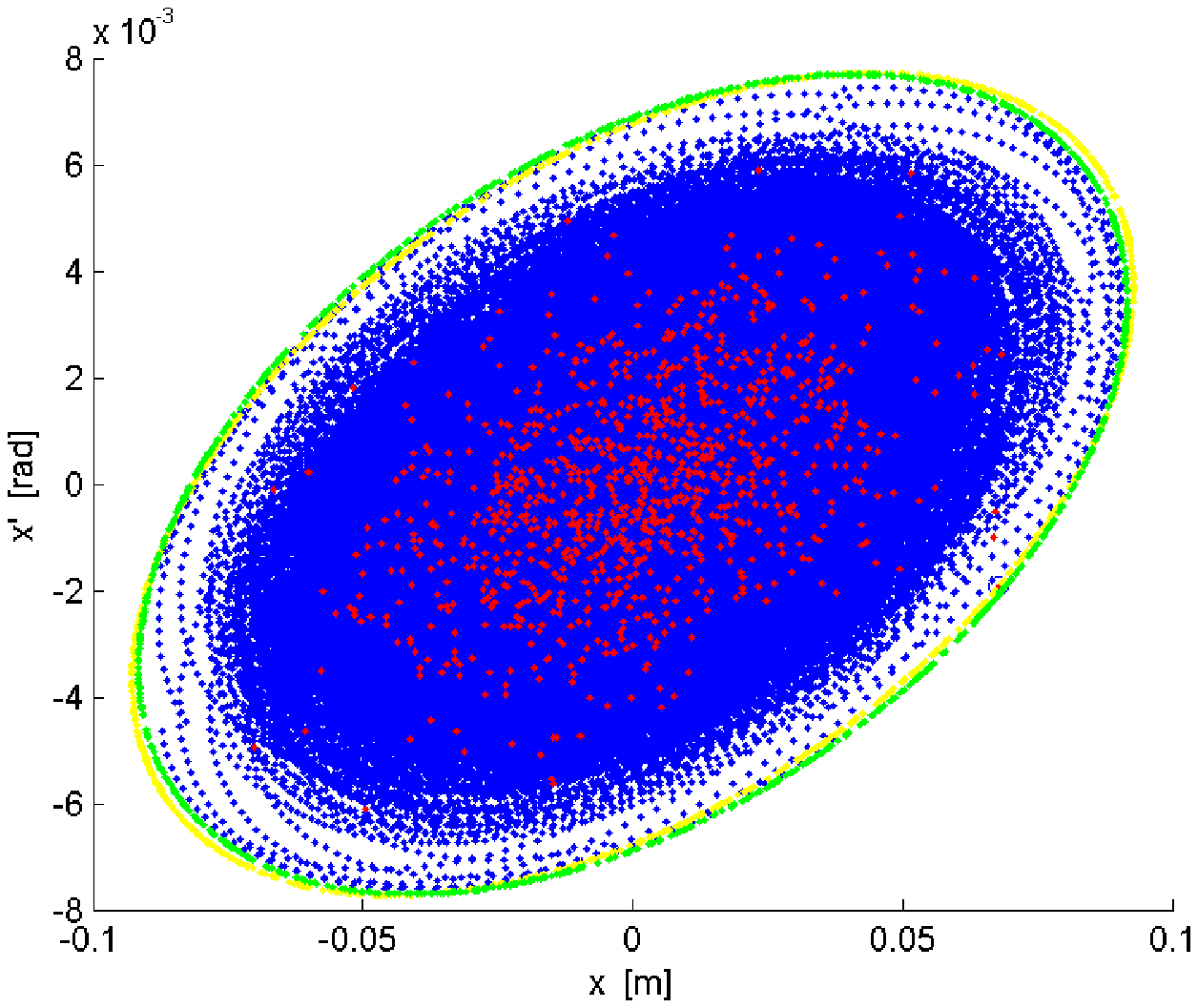}
\figcaption{\label{fig12} Projection of 1000 particles for 2000 turns in x direction. The yellow ellipse is the matched emittance cycle for initial beam. Red points are the initial beam. The green ellipse is the boundary cycle.}
\end{center}

From Figure 11, 12, one can found that the Footprints of the beam in x direction and in y direction are surrounded by the boundary ellipses perfectly. It verifies that the functions of the boundary ellipses obtained from Equation (2) are correct, and the coupled beam's boundary in phase plane can be expressed by the Riken's lattice function successfully.

In x direction (Figure 12), the boundary ellipse is very close to the matched emittance cycle of the initial beam. But in y direction (Figure 12), the boundary ellipse is much larger than the matched emittance cycle of the initial beam. This is due to the initial particles' horizontal emittance (200$\pi$ mm mrad) is much larger than the vertical one (30$\pi$ mm mrad).

From now, the coupled motion for a single particle and for a beam in CSRm are clearly expressed by Ripken's parameters. The function of projection's boundary in phase plane(Equation (6), (7)) is also obtained and parameterized by Riken's lattice function.

Comparing with beam's projection boundary in phase plane of the ring, one may care more about the envelops in the real space. So in next section a new method to calculate the coupled envelops will be shown.

\section{Coupled beam's envelope}

Assuming that at the beginning the initial particles are matched beam in the uncoupled ring (the solenoids are compensated by the compensation solenoids perfectly), and then the compensation solenoids are turned off suddenly. So after the compensation solenoids are turned off, the transverse motion will become coupled. The envelope of the beam will be different from the uncoupled case.

  Form Equation (2), it can be found that: for a particle the maximum value of x is equal to  $\sqrt {{{\rm{\varepsilon }}_1}{{\rm{\beta }}_{{\rm{x}}1}}}  + \sqrt {{{\rm{\varepsilon }}_2}{{\rm{\beta }}_{{\rm{x}}2}}} $, the maximum value of y is equal to $\sqrt {{{\rm{\varepsilon }}_1}{{\rm{\beta }}_{{\rm{y}}1}}}  + \sqrt {{{\rm{\varepsilon }}_2}{{\rm{\beta }}_{{\rm{y}}2}}} $ at a position of the ring.

   Hence to calculate the coupled envelope for a beam, the ${{{{\rm{\mathord{\buildrel{\lower3pt\hbox{$\scriptscriptstyle\frown$}}
\over \varepsilon } }}}_1}}$, ${{{{\rm{\mathord{\buildrel{\lower3pt\hbox{$\scriptscriptstyle\frown$}}
\over \varepsilon } }}}_2}}$  (particular values of $\epsilon_1$, $\epsilon_2$), with which the function $\sqrt {{{\rm{\varepsilon }}_1}{{\rm{\beta }}_{{\rm{x}}1}}}  + \sqrt {{{\rm{\varepsilon }}_2}{{\rm{\beta }}_{{\rm{x}}2}}} $ will reach the maximum, should be calculated out at every point of the ring. Fortunately, $\epsilon_1$, $\epsilon_2$ are constants for a particle when particle moves along ring [5], and the Ripken's lattice functions $\beta_{x1}$， $\beta_{x2}$ have the same monotonicity. So $\epsilon_1$, $\epsilon_2$ will have same values at different positions of the ring to make $\sqrt {{{\rm{\varepsilon }}_1}{{\rm{\beta }}_{{\rm{x}}1}}\left( {\rm{s}} \right)}  + \sqrt {{{\rm{\varepsilon }}_2}{{\rm{\beta }}_{{\rm{x}}2}}\left( {\rm{s}} \right)}$ to be maximized. So we just need to calculate  ${{{{\rm{\mathord{\buildrel{\lower3pt\hbox{$\scriptscriptstyle\frown$}}
\over \varepsilon } }}}_1}}$, ${{{{\rm{\mathord{\buildrel{\lower3pt\hbox{$\scriptscriptstyle\frown$}}
\over \varepsilon } }}}_2}}$ at one point , and then calculate the envelope of the beam by:
\begin{equation}\label{9}
  {E_x}\left( s \right) = \sqrt {{{{\rm{\mathord{\buildrel{\lower3pt\hbox{$\scriptscriptstyle\frown$}}
\over \varepsilon } }}}_1}{{\rm{\beta }}_{{\rm{x}}1}}\left( {\rm{s}} \right)}  + \sqrt {{{{\rm{\mathord{\buildrel{\lower3pt\hbox{$\scriptscriptstyle\frown$}}
\over \varepsilon } }}}_2}{{\rm{\beta }}_{{\rm{x}}2}}\left( {\rm{s}} \right)}, \
\end{equation}
\begin{equation}\label{10}
  {E_y}\left( s \right) = \sqrt {{{{\rm{\mathord{\buildrel{\lower3pt\hbox{$\scriptscriptstyle\frown$}}
\over \varepsilon } }}}_1}{{\rm{\beta }}_{y1}}\left( {\rm{s}} \right)}  + \sqrt {{{{\rm{\mathord{\buildrel{\lower3pt\hbox{$\scriptscriptstyle\frown$}}
\over \varepsilon } }}}_2}{{\rm{\beta }}_{y2}}\left( {\rm{s}} \right)} .\
\end{equation}

The ${{{{\rm{\mathord{\buildrel{\lower3pt\hbox{$\scriptscriptstyle\frown$}}
\over \varepsilon } }}}_1}}$, ${{{{\rm{\mathord{\buildrel{\lower3pt\hbox{$\scriptscriptstyle\frown$}}
\over \varepsilon } }}}_2}}$ are calculated by follows: The values of $\epsilon_1$, $\epsilon_2$ should satisfy equation (13), and at the same time they should let function $\sqrt {{{\rm{\varepsilon }}_1}{{\rm{\beta }}_{{\rm{x}}1}}}  + \sqrt {{{\rm{\varepsilon }}_2}{{\rm{\beta }}_{{\rm{x}}2}}} $ reach maximum.

\begin{equation}\label{11}
  \left\{ {\begin{array}{*{20}{c}}
{\sqrt {{{\rm{\varepsilon }}_1}{{\rm{\beta }}_{{\rm{x}}1}}} \cos \left( {{{\rm{\Phi }}_{{\rm{x}}1}} + {{\rm{\Phi }}_1}} \right) + \sqrt {{{\rm{\varepsilon }}_2}{{\rm{\beta }}_{{\rm{x}}2}}} \cos \left( {{{\rm{\Phi }}_{{\rm{x}}2}} + {{\rm{\Phi }}_2}} \right) = {{\rm{x}}_{\rm{\;}}}}\\
{\sqrt {{{\rm{\varepsilon }}_1}{{\rm{\gamma }}_{{\rm{x}}1}}} \cos \left( {{\phi _{{\rm{x}}1}} + {{\rm{\Phi }}_1}} \right) + \sqrt {{{\rm{\varepsilon }}_2}{{\rm{\gamma }}_{{\rm{x}}2}}} \cos \left( {{\phi _{{\rm{x}}2}} + {{\rm{\Phi }}_2}} \right) = {{{\rm{x'}}}_{\rm{\;}}}}\\
{\sqrt {{{\rm{\varepsilon }}_1}{{\rm{\beta }}_{{\rm{y}}1}}} \cos \left( {{{\rm{\Phi }}_{{\rm{y}}1}} + {{\rm{\Phi }}_1}} \right) + \sqrt {{{\rm{\varepsilon }}_2}{{\rm{\beta }}_{{\rm{y}}2}}} \cos \left( {{{\rm{\Phi }}_{{\rm{y}}2}} + {{\rm{\Phi }}_2}} \right) = {\rm{y}}}\\
\begin{array}{l}
\sqrt {{{\rm{\varepsilon }}_1}{{\rm{\gamma }}_{{\rm{y}}1}}} \cos \left( {{\phi _{{\rm{y}}1}} + {{\rm{\Phi }}_1}} \right) + \sqrt {{{\rm{\varepsilon }}_2}{{\rm{\gamma }}_{{\rm{y}}2}}} \cos \left( {{\phi _{{\rm{y}}2}} + {{\rm{\Phi }}_2}} \right) = {\rm{y'}}\\
{{\rm{\gamma }}_{\rm{x}}}{{\rm{x}}^2} + 2{{\rm{\alpha }}_{\rm{x}}}{\rm{x}}{{\rm{x}}^{\rm{'}}} + {{\rm{\beta }}_{\rm{x}}}{{{\rm{x'}}}^2} \le {{\rm{A}}_{\rm{x}}}\\
{{\rm{\gamma }}_{\rm{y}}}{{\rm{y}}^2} + 2{{\rm{\alpha }}_{\rm{y}}}{\rm{y}}{{\rm{y}}^{\rm{'}}} + {{\rm{\beta }}_{\rm{y}}}{{{\rm{y'}}}^2} \le {{\rm{A}}_{\rm{y}}}
\end{array}
\end{array}} \right..\
\end{equation}

Where the $\gamma_x$, $\gamma_y$, $\alpha_x$, $\alpha_y$, $\beta_x$, $\beta_y$  are uncoupled twiss parameters of a ring.

By numerical calculation, the ${{{{\rm{\mathord{\buildrel{\lower3pt\hbox{$\scriptscriptstyle\frown$}}
\over \varepsilon } }}}_1}}$, ${{{{\rm{\mathord{\buildrel{\lower3pt\hbox{$\scriptscriptstyle\frown$}}
\over \varepsilon } }}}_2}}$ are obtained for CSRm: ${{{{\rm{\mathord{\buildrel{\lower3pt\hbox{$\scriptscriptstyle\frown$}}
\over \varepsilon } }}}_1}}$ = 0.00063964077389 and ${{{{\rm{\mathord{\buildrel{\lower3pt\hbox{$\scriptscriptstyle\frown$}}
\over \varepsilon } }}}_2}}$ = 0.00009085803501.

\begin{center}
\includegraphics[width=6.5cm]{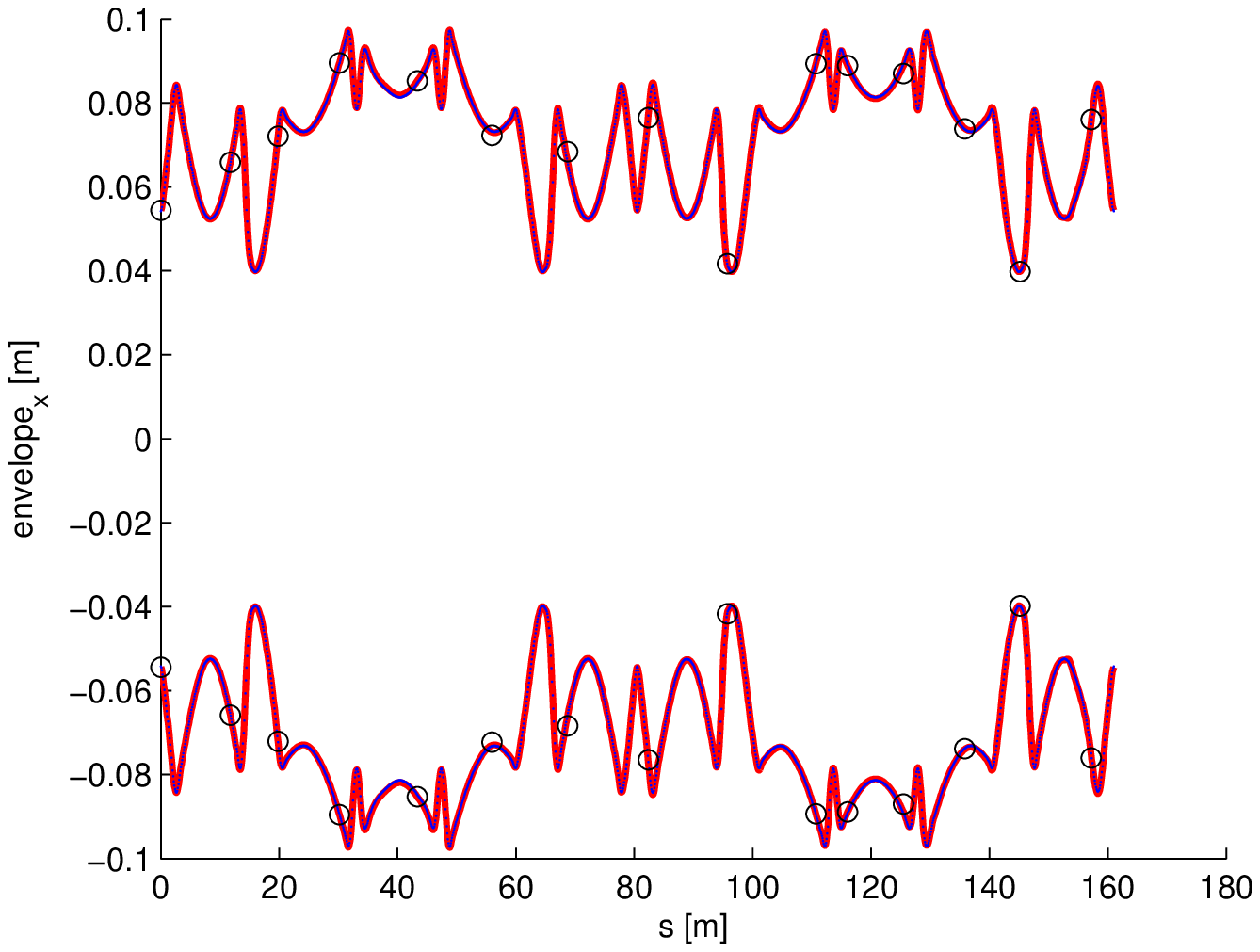}
\figcaption{\label{fig13} The coupled envelope in x direction.}
\end{center}

Using equation (11), (12) and the Ripken's lattice functions $\beta_{x1}$, $\beta_{x2}$, $\beta_{y1}$, $\beta_{y2}$ of the ring CSRm, the coupled envelope of the beam can be calculated out. Figure (13), (14) show the envelopes in x direction and y direction. To check the reliability of our method, the results obtained by beam tracking are also shown in Figure 13, 14.

\begin{center}
\includegraphics[width=6.5cm]{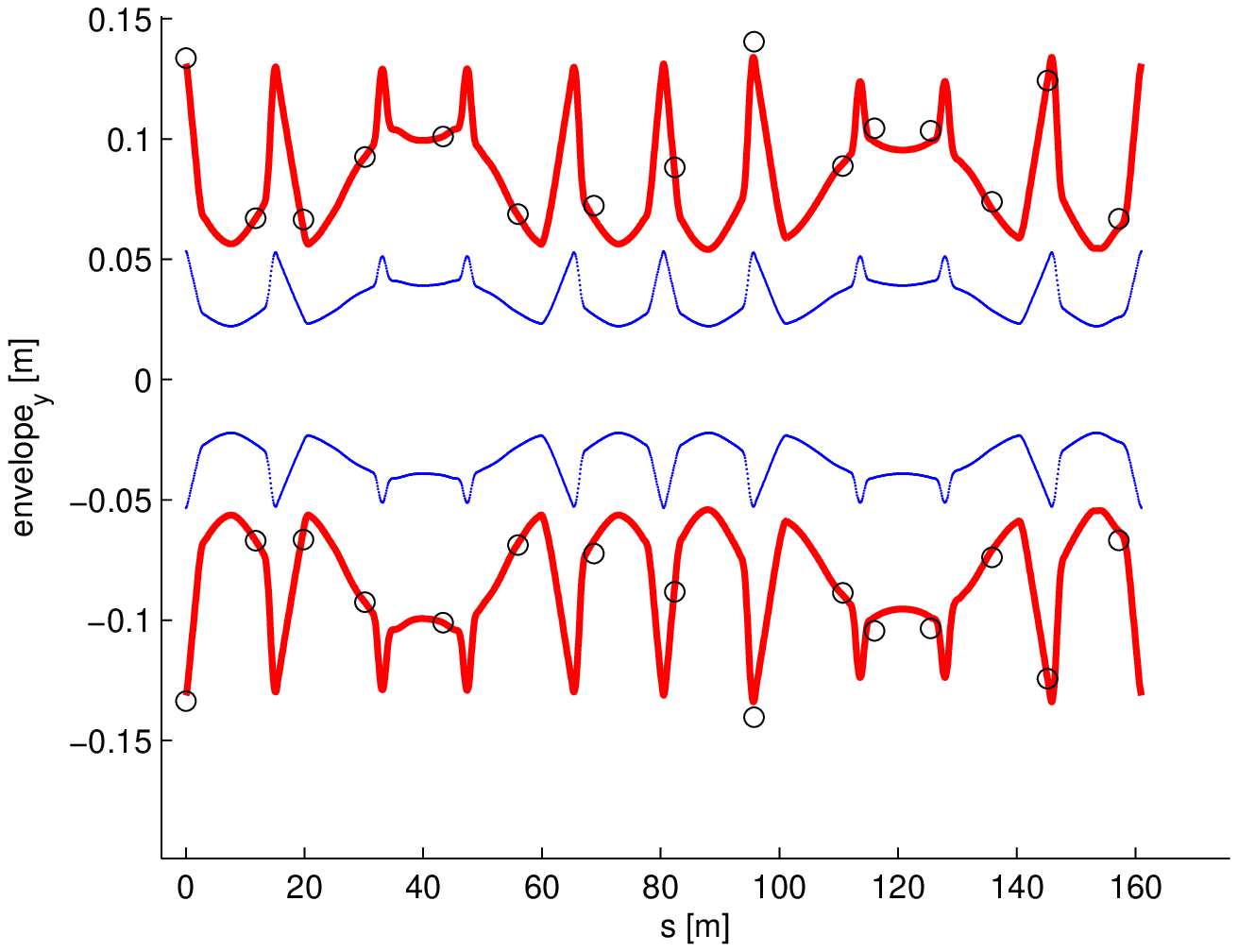}
\figcaption{\label{fig14} The coupled envelope in y direction.}
\end{center}

The coupled envelopes of the CSRm calculated by the method mentioned above are presented in Figure13, 14 by the red lines, and the envelopes in some positions of the ring obtained by tracking are shown by the cycle marker. It can be found that the tracking results of the envelope are accordance with the results calculated by numerical method, it verifies that the new method used to calculated the coupled beam envelopes is reliable.

As a compare, uncoupled envelope of the beam also be presented by blue line. From Figure 13, it can be found that in x direction the uncoupled and coupled envelopes are nearly have the same value, the blue line are nearly covered by the red line. From Figure 14, one can find that in y direction the coupled envelope is much larger than the uncoupled one.

The coupled beam envelope is exceeding the dimension of aperture ($ \pm $30mm in dipole, $ \pm $50mm in quadruple )[1] in y direction. So if there is no compensation solenoid, the coupling induced by cooler's solenoids will cause significant beam loss, and most particles lost in y direction.

\section{Conclusions}

The coupled transverse motion induced by cooler's solenoids of CSRm is calculated. To explain the simulation results the Ripken's approach that used to parameterize the coupled betatron motion is introduced. The Ripken's lattice functions are calculated for CSRm. The functions of boundary ellipse in phase plane for coupled beam are obtained. The coupled envelopes of CSRm are calculated by a new method. The conclusions are: 1. If there is no compensation solenoid, the coupling induced by cooler's solenoids will cause significant beam loss, and most particles lost in y direction. 2. The new method used to calculated the coupled envelopes are available.

\vspace{-1mm}
\centerline{\rule{30mm}{0.4pt}}

\end{multicols}

\clearpage

\end{document}